\documentclass[]{hdsr}

\graphicspath{{figs/}}

\begin{document}

\newgeometry{bottom=1.5in}

\volumeheader{0}{0}{00.000}

\begin{center}

  \title{Translation Readiness Index: Measuring the Semantic Proximity of Research to Patented Science}
  \maketitle

  \thispagestyle{empty}
  
  \vspace*{.2in}

  \begin{tabular}{cc}
    Paul X. McCarthy\upstairs{\affilone\affiltwo\affilthree,*}, Rasika Amarasiri\upstairs{\affilone}, Xian Gong\upstairs{\affilone}
   \\[0.25ex]
   {\small \upstairs{\affilone} League of Scholars, Sydney, 1225, NSW, Australia,} \\
   {\small \upstairs{\affiltwo} UNSW Sydney, Sydney, 2052, NSW, Australia,} \\
   {\small \upstairs{\affilthree} University of Technology Sydney, Sydney, 2007, NSW, Australia} \\
  \end{tabular}
  
  \emails{
    \upstairs{*}paul@leagueofscholars.com
    }
  \vspace*{0.4in}

\begin{abstract}
Universities, funders, and investors often need to spot research with translational potential early, long before downstream outcomes like licenses, startups, or patents emerge. We introduce the Translation Readiness Index (TRI), a scalable text-based metric that estimates a publication's semantic proximity to patent-linked science using only its title and abstract. Trained on over 20,000 scientific papers, contrasting papers paired with U.S. patents for the same invention against non-patent papers from the same journals, TRI uses domain-specific document embeddings to detect latent linguistic signals. Patent-paired papers consistently use an action-oriented “language of invention”, whereas non-patent papers favor observational framing. Using only titles and abstracts, the model accurately distinguishes patent-paired research from comparison papers (ROC-AUC = 0.774). External validation across independent datasets shows that higher TRI scores strongly align with real-world translational activity. High scores correlate with industry coauthorship, author patent histories, and independent commercial-potential benchmarks. At the institutional level across leading global universities, average TRI correlates significantly with university-industry collaboration ($r = 0.364, p < 0.001$). TRI provides an automated early-stage screening tool to prioritize research for expert review, measuring semantic proximity to patented science rather than guaranteed commercial outcomes.
\end{abstract}
\end{center}

\vspace*{0.15in}
\hspace{10pt}
  
\copyrightnotice

\section*{Media Summary}
Public research is funded in part on the expectation that some of it will feed into new technology, industry, health, and policy. But whether a piece of research is heading in that direction usually only becomes clear years later, once patents are filed, licenses are signed, or startups are formed. That delay makes it hard for universities, funders, and investors to spot promising work early, when their decisions matter most.

This study introduces the Translation Readiness Index (TRI), a score that reads only a paper's title and abstract and estimates how closely it resembles research that has historically underpinned patented inventions. The idea rests on a simple observation: there appears to be a detectable “language of invention”. Papers tied to patents tend to talk about designing, developing, and building things, such as prototypes, devices, techniques, while otherwise similar papers lean toward observing, explaining, and describing.

To test this, we drew on the first open, all-field database of patent-paper pairs, roughly half a million pairs built by matching U.S. patents to scientific papers. We compared several ways of turning text into numbers, from simple word counts to modern AI language models trained on scientific writing, and found the language signal is real and measurable. We then checked the score against independent evidence: papers with higher TRI scores were more likely to be written by inventors, to involve “use-inspired” researchers who both publish and patent, and to be coauthored with industry. Universities with higher average TRI tended to collaborate more with industry.

TRI is not a prediction that a paper will be commercialized, and it should never be used on its own to decide funding, hiring, or promotion. It can undervalue fields where impact flows through clinical guidelines, public policy, or open-source software rather than patents. What it offers is an early, scalable, text-based signal, a way to help people screen and prioritize research for closer expert review, long before the usual indicators appear.

\section{Introduction}
\label{sec1}
Scientific publications are one route through which research enters technological development. Because of this link, public research funding is often justified in part by the expectation that some research will contribute to technology, industry, health, policy, or other public outcomes \citep{jaffe1989real,nelson1993national,furman2002determinants}. Yet, despite these expectations, most publications do not have a direct or observable path into patents, licensing, startups, industry use, policy, or practice \citep{mowery2004ivory,perkmann2013academic,fortunato2018science}. This disconnect creates a challenge for universities, funders, investors, and corporate research groups, who often need earlier indicators of a publication's translational relevance \citep{cohen2002links,veugelers2014contribution,ahmadpoor2017dual}. Establishing these earlier indicators could actively support funding decisions, technology-transfer reviews, and expert screening of publications \citep{narin1997increasing,marx2020reliance}.

Prior work has extensively explored links between scientific publications and technological invention. Early studies demonstrated that patents frequently cite scientific literature, providing foundational evidence of this relationship \citep{narin1997increasing}. Subsequent research established patent citations to scientific literature as an important mechanism for measuring knowledge flows from science to technology \citep{jaffe1993geographic,henderson1998universities}. More recently, studies of patent-paper pairs have revealed deep connections between academic research and invention in fields such as biotechnology, pharmaceuticals, information technology, and advanced manufacturing \citep{cohen2002links,fleming2004science,marx2020reliance}. Today, the availability of large bibliometric and patent databases allows researchers to study links between publications, patents, inventors, institutions, and firms at an unprecedented scale \citep{fortunato2018science,wang2021science}.

Existing indicators, however, suffer from key limitations. Many indicators are observed after publication, including patents, licenses, startup formation, industry funding, and commercialization revenue \citep{mowery2004ivory,shane2004academic,markman2005entrepreneurship}. Patent citations provide evidence that a publication is used in patenting activity, but they are observed only after patent documents are filed and processed \citep{trajtenberg1990penny,harhoff2003citations}. Institutional metrics such as licensing income and startup formation also depend on organizational capacity, technology-transfer policy, and field \citep{siegel2003commercial,geuna2006university,perkmann2013academic}. These measures therefore provide limited ability to identify translational opportunities at the publication stage when strategic decisions regarding funding, collaboration, and commercialization pathways are most valuable \citep{veugelers2014contribution,fortunato2018science}.

A distinct and tighter signal comes from patent-paper pairs (PPPs). In a patent-paper pair, a scientific paper is linked to a patent that protects the same underlying invention, typically because the paper and patent were produced by the same individuals; many, though not all, such pairs also involve the patent citing the paper. This is a much tighter relationship than a patent simply citing a paper as background. Patent-paper pairs have been used in innovation and science-of-science research for over two decades to trace how scientific discoveries become commercial inventions, but progress has long been constrained by data: prior PPP datasets were generally built study-by-study, were mostly confined to the life sciences, and were not openly available. We draw on the first open, all-field database of such pairs \citep{marx2024does}: roughly half a million pairs spanning 1976 to 2023, constructed by linking U.S.\ patents from PatentsView to scientific papers in OpenAlex and graded into four confidence tiers. This resource makes it possible, for the first time, to study the language of patent-linked science across all fields at scale.

Scientific document embeddings allow publications to be compared using title and abstract text \citep{devlin2018bert,bommasani2021foundation}. Scientific document embeddings are created by encoding title and abstract text as vectors that represent each publication in the context of a large linguistic corpus of over ten million other peer-reviewed research publications and their citation networks that then can be used in classification or similarity tasks \citep{cohan2020specter,beltagy2019scibert,singh2023specter2}. If publications in patent-paper pairs differ in language from matched comparison publications, a classifier based on contextual embeddings using title and abstract data alone should be able to detect that difference \citep{porter2005tech,arora2013capturing}. If successful, such an approach would then allow a translational-orientation score to be inferred for any published paper.

We introduce the Translation Readiness Index (TRI), a measure of similarity to publications that appear in high-confidence patent-paper pairs. The method uses more than 20,000 scientific publications labeled using patent-paper-pair data \citep{marx2024does} drawn from the open all-field PPP database described above. To establish the best way to capture the language signal, we systematically compared five text representations, including bag-of-words, TF-IDF, a general-purpose sentence encoder (MiniLM), SciBERT, and SPECTER2 scientific document embeddings, each crossed with multiple classifiers and evaluated by held-out ROC-AUC. We then define TRI using the SPECTER2 embedding \citep{singh2023specter2} with an XGBoost classifier \citep{chen2016xgboost}. Although SPECTER2 with XGBoost is not the single highest-scoring combination, it is selected on measurement grounds rather than leaderboard position: SPECTER2 is a document-level representation designed specifically for scientific papers, is a de facto standard in science-of-science research, and can be computed at scale for any paper with a title and abstract, which matters for an index intended to score millions of publications reproducibly. The resulting prediction probabilities are used as TRI scores. A higher score means that the publication is more similar to publications in the patent-paper-paired class.

\begin{figure}[ht]
  \centering
  \includegraphics[width=\textwidth]{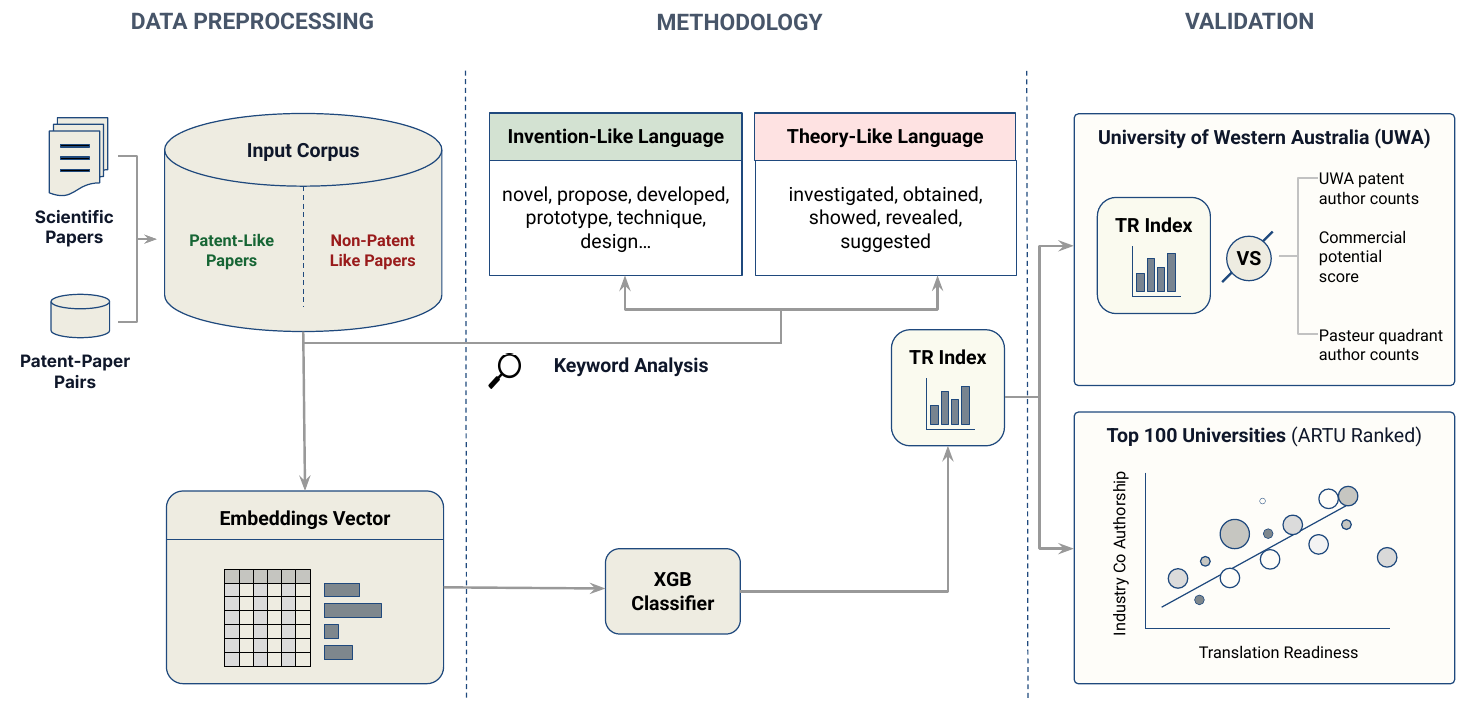}
  \vspace{0.1pt}
  \caption{\textbf{Overview of the Translation Readiness Index (TRI) framework.} Scientific publications are divided into patent-paper-paired and matched comparison groups. Titles and abstracts are encoded using SPECTER2 embeddings. An XGBoost classifier estimates the probability that each publication belongs to the patent-paper-paired class. This probability is defined as the Translation Readiness Index. Keyword analysis is used to interpret language differences between the two groups; it is not an input to the TRI model. The resulting index is validated against established indicators of research translation, including patenting activity, commercial potential, Stokes' Pasteur's Quadrant participation, and industry coauthorship across leading universities.}
  \label{fig:fig1}
\end{figure}

This study has three parts. First, we develop a text-based measure to assess whether a publication resembles scientific papers in high-confidence patent-paper pairs. Second, we test whether semantic embeddings can distinguish patent-paper-paired publications from matched comparison publications using title and abstract text, and we benchmark this signal across five text representations. Third, we compare TRI scores with independent publication-, author-, and institution-level indicators related to research translation, including patent authorship, Stokes' Pasteur's Quadrant researchers, and industry coauthorship. We also compare TRI scores with commercial-potential scores derived using an independent inference method and independent data. The study provides a text-based measure that can be compared with other indicators of research translation. The workflow of the TRI is shown in Figure~\ref{fig:fig1}.

\section{Research Background}

\subsection{Scientific Papers and Patent Data}
Science and innovation studies have long examined links between scientific discovery and technological invention \citep{nelson1959simple,jaffe1989real,kline2009overview}. Since Bush \citep{bush1960science}, policy discussions have often treated scientific research as an upstream input to technological progress and economic growth. Early empirical work used patent citations to study links between scientific publications and invention \citep{griliches1990patent,jaffe1993geographic}.

An important step came from \citet{narin1997increasing}, who showed that patent documents increasingly cited scientific literature. Subsequent studies found strong connections between academic publications and patenting activity in biotechnology, pharmaceuticals, semiconductors, and advanced materials \citep{cohen2002links,henderson1998universities,fleming2004science}. Patent citations to scientific articles became common indicators of technological relevance \citep{trajtenberg1990penny,harhoff2003citations,ahmadpoor2017dual}.

Beyond citations, a smaller body of work has studied patent-paper pairs \citep{magerman2015does}, in which a scientific paper and a patent are matched because they describe the same underlying invention, typically with overlapping authors and inventors. These pairs provide a tighter science-technology link than background citation. Historically, however, this literature faced a practical constraint: each study constructed its own set of pairs, coverage was concentrated in the life sciences, and the resulting datasets were rarely shared. The recent construction of the first open, all-field database of patent-paper pairs \citep{marx2024does}, roughly half a million pairs from 1976 to 2023, built by matching U.S. patents from PatentsView to OpenAlex publications, removes much of this constraint and enables analysis of patent-linked science across all disciplines. Our study builds directly on this resource.

Large bibliographic and patent databases expanded work in this area. Databases such as Web of Science, Scopus, OpenAlex, Microsoft Academic Graph, Google Patents, Lens.org, and PATSTAT allow researchers to link publications, patents, inventors, institutions, and firms \citep{fortunato2018science}. These studies found that papers cited by patents often receive more citations and more industrial attention than papers without patent connections \citep{narin1997increasing,ahmadpoor2017dual,marx2020reliance}.

Recent research has expanded beyond citation analysis to examine direct author-level and institution-level relationships between scientific research and technological invention. Studies of academic inventors show that some researchers publish and patent in related areas, often working across universities, firms, hospitals, or research institutes \citep{balconi2004networks,colyvas2002university,perkmann2013academic}. Research on Stokes' Pasteur's Quadrant describes work that combines basic research questions with practical use \citep{stokes1997pasteur,grant2003basic}.

Machine-learning methods are now being used to study relationships between scientific publications and patents. Researchers increasingly use text mining, semantic embeddings, topic modeling, and language models to compare publication and patent text \citep{porter2005tech,arora2013capturing,fortunato2018science}. Embedding-based approaches have been used to recover scientific relationships, predict citations and identify emerging technologies \citep{cohan2020specter,singh2023specter2,gong2025cosmos}. These studies suggest that the publications themselves may contain latent linguistic information, a “language of invention” capable of distinguishing applied, patent-like research from other more abstract or theoretical research.

\subsection{Current Measures of Research Commercialization Potential}
Research commercialization is often studied using indicators observed after publication \citep{griliches1990patent,mowery2004ivory,perkmann2013academic,masclans2025compot}. Patents are widely used because they record legally protected inventions \citep{trajtenberg1990penny,harhoff2003citations}. Prior studies report associations between patenting activity and economic outcomes such as firm growth, venture capital investment, and innovation performance \citep{hall2005market,ahmadpoor2017dual,arora2023effect}.

Technology-transfer offices often track licensing revenue, invention disclosures, startup formation, and industry-sponsored research \citep{colyvas2002university,siegel2003commercial,markman2005entrepreneurship}. These measures are useful because they record observable outcomes \citep{mowery2004ivory,shane2004academic}. However, they vary across disciplines and institutions \citep{geuna2006university,perkmann2013academic}. Some fields produce clinical, environmental, policy, educational, or cultural value without producing many patents or licenses \citep{bornmann2013societal,wouters2019rethinking}.

Other approaches address parts of this problem. Stokes’ Pasteur’s Quadrant framework distinguishes between pure basic research, pure applied research, and use-inspired basic research \citep{stokes1997pasteur}. This framework is now widely used to evaluate translational research by identifying 'patenting scholars' and tracing how fundamental science yields clinical and commercial outputs \citep{grant2003basic,veugelers2014contribution}. Publication-level measures include patent citation counts, industry citation measures, technology relevance scores, and commercial-potential indicators \citep{narin1997increasing,ahmadpoor2017dual,marx2020reliance}.

Current measures have three limits \citep{perkmann2013academic,fortunato2018science}. First, many indicators are observed only after patenting, licensing, startup formation, or collaboration \citep{trajtenberg1990penny,mowery2004ivory}. Second, many measures depend on institutional factors such as technology-transfer capacity, patenting culture, and funding availability \citep{siegel2003commercial,geuna2006university}. Third, fields differ in how research enters patents, products, policy, clinical practice, software, or standards \citep{bornmann2013societal,veugelers2014contribution}.

These limits motivate a publication-level predictive measure that is based on the publication itself rather than later institutional outcomes. Instead of relying only on later outcomes, publication text and its context in a large model of over 10 million others can be used to estimate translational orientation at the publication level \citep{porter2005tech,arora2013capturing,fortunato2018science}. Such a measure could help identify publications with the highest translation potential before downstream indicators are available \citep{wang2021science}.

\section{Materials and Methods}
This study uses patent-paper pairs as the positive training signal. A patent-paper pair identifies a scientific publication whose content has been matched to a patent document with high confidence. The positive class is therefore not simply papers that are cited in patents but papers that are very similar to patents themselves. This pairing does not, however, serve as direct evidence that a publication has produced licensing income, a startup, or a commercial product. In this paper, patent-paper-paired publications are used as an operational proxy for research that has direct representation in the patent system.

\subsection{Data}
The Translation Readiness Index framework uses publication metadata, patent-paper-pair data, researcher-level indicators, institutional publication data, and external validation measures, summarized in Table~\ref{table:tab1}. The training corpus is used to distinguish patent-paper-paired publications from matched comparison publications. The validation datasets are used to assess whether higher TRI scores are aligned with other independent indicators related to research translation.

The primary training corpus consisted of 20,610 peer-reviewed publications retrieved from OpenAlex, including 9,431 patent-like and 11,179 non-patent-like papers. Positive samples were constructed using the Reliance on Science (RoS) patent-paper-pair data \citep{marx2024does}. These data identify scientific publications matched to patent documents. The version used here reports confidence on a four-level scale. To ensure high-quality labels, only records with the highest confidence (levels 3 or 4) were retained, comprising 188,358 valid papers in OpenAlex. Confidence levels 3 and 4 denote high-confidence patent-paper matches supported by strong connection evidence, and are therefore commonly used to maximize the precision and reliability of identified science–technology connections. From these records, 9,431 publications from 2019 to 2026 with available title and abstract text were used as positive examples. These publications are described as patent-paper-paired publications. They are not treated as evidence of realized commercialization.

Matched comparison publications (n = 11,179) were selected from OpenAlex after excluding publications appearing in the patent-paper-pair corpus. Comparison publications were sampled to match the positive corpus by publication year, journal venue, and OpenAlex subfield where possible. This design reduces, but does not eliminate, confounding by time, outlet, and discipline. The resulting dataset provides the supervised classification corpus.

For external validation, we used 25,921 publications authored by researchers affiliated with the University of Western Australia (UWA) between 2019 and 2026. Additional validation datasets include 1,254 UWA-affiliated patent authors identified through Google Patents, 694 UWA publications with commercial-potential scores from the ComPot framework \citep{masclans2025compot}, and 6,546 UWA researchers classified in Stokes' Pasteur’s Quadrant Research Score (PQRS) data. PQRS identifies researchers who both publish and patent, constructed by linking disambiguated OpenAlex author profiles to disambiguated PatentsView inventors over the period 1976–2023. Institutional-level validation uses publication data from the top 100 universities in the Aggregate Ranking of Top Universities (ARTU) and industry collaboration indicators from the CWTS Leiden Ranking. These datasets allow examination of the relationship between TRI scores and independent measures of university-industry engagement. Together they support publication-level, author-level, and institution-level comparisons between TRI and external indicators related to research translation.

\begin{table*}[ht]
\centering
\caption{\textbf{Data sources used in the Translation Readiness Index study.} The study uses publication, patent-paper pairs, researcher, institutional, and commercial-potential datasets to construct and validate TRI.}
  \label{table:tab1}
  \vspace{0.3cm}
  \small
  \makebox[\textwidth][c]{%
  \hspace*{-0.3cm}
  \begin{tabular}{cccc}
    \toprule
    \textbf{Component} & \textbf{Data Source} & \textbf{Description} & \textbf{Sample Size}\\
    \midrule
    \multicolumn{1}{c}{\makecell{Key Training Data \\ (Peer-reviewed Papers)}} & \multicolumn{1}{c}{\makecell{OpenAlex}}  & \multicolumn{1}{c}{\parbox{7cm}{Balanced patent-like and non-patent-like papers with metadata.}} & \multicolumn{1}{c}{\makecell{20,610}} \\
    \midrule[0.3pt]
    \multicolumn{1}{c}{\makecell{Patent-Paper Pairs}} & \multicolumn{1}{c}{\makecell{Reliance on Science (PPP)}}  & \multicolumn{1}{c}{\parbox{7cm}{Scientific papers related to registered patents through high-confidence matching.}} & \multicolumn{1}{c}{\makecell{9,431}} \\
    \midrule[0.3pt]
    \multicolumn{1}{c}{\makecell{UWA Papers}} & \multicolumn{1}{c}{\makecell{OpenAlex}}  & \multicolumn{1}{c}{\parbox{7cm}{Peer-reviewed publications coauthored by researchers affiliated with the University of Western Australia (2019–2026).}} & \multicolumn{1}{c}{\makecell{25,921}} \\
    \midrule[0.3pt]
    \multicolumn{1}{c}{\makecell{UWA Patent Authors}} & \multicolumn{1}{c}{\makecell{Google Patents}}  & \multicolumn{1}{c}{\parbox{7cm}{UWA-affiliated inventors identified from patents across all jurisdictions.}} & \multicolumn{1}{c}{\makecell{1,254}} \\
    \midrule[0.3pt]
    \multicolumn{1}{c}{\makecell{Commercial Potential \\ Scores}} & \multicolumn{1}{c}{\makecell{ComPot}}  & \multicolumn{1}{c}{\parbox{7cm}{Publication-level commercial potential scores derived from the ComPot framework.}} & \multicolumn{1}{c}{\makecell{694}} \\
    \midrule[0.3pt]
    \multicolumn{1}{c}{\makecell{Pasteur's Quadrant \\ Research Score}} & \multicolumn{1}{c}{\makecell{PQRS}}  & \multicolumn{1}{c}{\parbox{7cm}{Researchers with both publication and patent records, representing use-inspired research activity.}} & \multicolumn{1}{c}{\makecell{6,546}} \\
    \midrule[0.3pt]
    \multicolumn{1}{c}{\makecell{Aggregate Ranking of \\ Top Universities}} & \multicolumn{1}{c}{\makecell{ARTU}}  & \multicolumn{1}{c}{\parbox{7cm}{Publication data for universities included in the ARTU Top 100, which combines ARWU, QS, and THE rankings.}} & \multicolumn{1}{c}{\makecell{8,111}} \\
    \midrule[0.3pt]
    \multicolumn{1}{c}{\makecell{Industry collaboration \\ rankings}} & \multicolumn{1}{c}{\makecell{CWTS\\Leiden Ranking}}  & \multicolumn{1}{c}{\parbox{7cm}{University-level measure of publications coauthored with one or more industrial partners.}} & \multicolumn{1}{c}{\makecell{1,594}} \\
  \bottomrule
  \end{tabular}}
\end{table*}

\subsection{Linguistic Analysis}
To compare the language of the two classes, we analyzed titles and abstracts from patent-paper-paired and matched comparison publications. The analysis used the combined title and abstract text for each publication.

Titles and abstracts were concatenated into a single text field before preprocessing. All text was converted to lowercase, URLs were removed, and excessive whitespace was normalized. Punctuation and domain-specific terms were largely preserved because technical expressions may carry class information. Documents lacking titles, abstracts, or patent-likeness labels were excluded from the analysis.

The cleaned corpus was divided into patent-paper-paired and matched comparison groups using the binary labels derived from the patent-paper pairs data. To compare terminology across groups, we used frequency-based $n$-gram analysis with Scikit-learn’s \texttt{CountVectorizer} \citep{pedregosa2011scikit}. Unigrams, bigrams, and trigrams were extracted ($n$-gram range $= 1$--$3$), while English stopwords were removed. To remove rare terms and very common phrases, only terms appearing in at least five documents were retained (\texttt{min\_df} = $5$), and terms appearing in more than 80\% of documents were excluded (\texttt{max\_df} = $0.8$).

To identify class-distinctive terms, we used term frequency-inverse document frequency (TF-IDF) \citep{salton1988term}. Using the same vocabulary construction parameters as the frequency analysis, TF-IDF weights were computed across the entire corpus. For each phrase, mean TF-IDF scores were calculated separately for patent-like and non-patent-like publications. The difference between these class-specific means was subsequently used to identify terms that contributed disproportionately to each document category. TF-IDF gives more weight to terms that are common in one group but less common across the corpus.

The highest-ranking phrases from the frequency and TF-IDF analyses were manually grouped into thematic categories. We identified recurring categories, including invention-oriented language, engineered systems and devices, signal-processing terms, applied biotechnology terms, observational language, and theoretical research framing. These categories are used in the Results section to interpret language differences between patent-paper-paired and matched comparison publications. The keyword analysis is used only for interpretation, which is not an input to the TRI model.

\subsection{Translation Readiness Index (TRI)}
TRI uses scientific document embeddings and supervised classification. The objective is to estimate whether a publication resembles other publications in high-confidence patent-paper pairs based on document-level embeddings encoding its title, abstract, and contextual position in a global citation network of over 10 million other papers.

\subsubsection{Model comparison}
TRI is built from two components: a \emph{text representation} that converts a paper's title and abstract into a numeric vector, and a \emph{classifier} that maps that vector to a probability of belonging to the patent-paper-paired class. Because the quality of the index depends on both components, we selected each component empirically by comparing candidates on a common 70:30 stratified training/test split and ranking them by held-out ROC-AUC. The candidate models for both components are summarized in Table~\ref{table:tab2}, and the full comparison is reported in Section~\ref{baseline_models}.

\begin{table*}[ht]
\centering
\caption{\textbf{Model Comparison.} The two components of TRI are: (a) text representation, which converts a paper's title and abstract into a numeric vector; and (b) classification, which maps that vector to a patent-paper-paired probability, and the candidate models are compared for each.}
  \label{table:tab2}
  \vspace{0.3cm}
  \small
  \makebox[\textwidth][c]{%
  \hspace*{-0.3cm}
  \begin{tabular}{ccl}
    \toprule
    &\textbf{Models}&\textbf{Description}\\
    \midrule
    \multirow{5}{*}{\rotatebox[origin=c]{90}{\parbox[c]{4.5cm}{\centering \textbf{Text Representation}}}} & \multicolumn{1}{c}{\makecell{\textbf{Bag-of-words (BOW)}}}  & \multicolumn{1}{c}{\parbox{12cm}{Sparse vector of raw term counts; no weighting and no word order or context.}}  \\
    \cmidrule(lr){2-3} 
    & \multicolumn{1}{c}{\makecell{\textbf{TF-IDF}}}  & \multicolumn{1}{c}{\parbox{12cm}{Term-frequency/inverse-document-frequency weighted counts, up-weighting terms distinctive to a document and down-weighting corpus-wide ones.}} \\
    \cmidrule(lr){2-3} 
    & \multicolumn{1}{c}{\makecell{\textbf{MiniLM}}}  & \multicolumn{1}{c}{\parbox{12cm}{A compact general-purpose sentence encoder producing dense embeddings; not adapted to scientific writing.}} \\
    \cmidrule(lr){2-3} 
    & \multicolumn{1}{c}{\makecell{\textbf{SciBERT}}}  & \multicolumn{1}{c}{\parbox{12cm}{A BERT-family transformer pretrained on a large scientific corpus, yielding context-aware embeddings tuned to scientific vocabulary.}} \\
    \cmidrule(lr){2-3} 
    & \multicolumn{1}{c}{\makecell{\textbf{SPECTER2}}}  & \multicolumn{1}{c}{\parbox{12cm}{A citation-aware, document-level embedding model for scientific papers; encodes the title and abstract into a 768-dimensional vector informed by the citation graph.}} \\

    \midrule
    \multirow{5}{*}{\rotatebox[origin=c]{90}{\parbox[c]{4cm}{\centering \textbf{Classification}}}} & \multicolumn{1}{c}{\makecell{\textbf{Naïve Bayes}}}  & \multicolumn{1}{c}{\parbox{12cm}{Probabilistic baseline applying Bayes' theorem under a feature-independence assumption; fast and effective on sparse count features.}} \\
    \cmidrule(lr){2-3} 
    & \multicolumn{1}{c}{\makecell{\textbf{Logistic Regression}}}  & \multicolumn{1}{c}{\parbox{12cm}{Linear model estimating class probability through the logistic function; an interpretable baseline for both sparse and dense inputs.}} \\
    \cmidrule(lr){2-3} 
    & \multicolumn{1}{c}{\makecell{\textbf{Random Forest}}}  & \multicolumn{1}{c}{\parbox{12cm}{Ensemble of decision trees trained on bootstrapped samples with feature subsampling; predictions averaged to reduce variance.}} \\
    \cmidrule(lr){2-3} 
    & \multicolumn{1}{c}{\makecell{\textbf{Gradient Boosting}}}  & \multicolumn{1}{c}{\parbox{12cm}{Ensemble that fits decision trees sequentially, each correcting the residual errors of the previous ones, capturing nonlinear structure.}} \\
    \cmidrule(lr){2-3} 
    & \multicolumn{1}{c}{\makecell{\textbf{Extreme Gradient Boosting} \\ \textbf{(XGBoost)}}}  & \multicolumn{1}{c}{\parbox{12cm}{Regularized, scalable implementation of gradient-boosted trees with efficient training.}} \\
  \bottomrule
  \end{tabular}}
\end{table*}

\textbf{Text representation.} 
The five candidate text representations span a progression from sparse lexical representations to dense, citation-aware scientific document embeddings (Table~\ref{table:tab2}, top). Bag-of-words (BOW) represents each publication as a sparse vector of raw term frequencies over the corpus vocabulary, treating each term independently and disregarding word order and semantic context, thereby providing a simple lexical baseline \citep{harris1954distributional}. TF-IDF extends this representation by weighting terms according to their frequency within a document and their rarity across the corpus, thereby emphasizing vocabulary that is more discriminative for individual publications \citep{salton1988term}. MiniLM is a compact, general-purpose transformer-based sentence encoder that produces dense semantic embeddings from large-scale language pretraining. Because it is trained on general-domain text rather than scientific literature, it provides a domain-agnostic dense embedding baseline \citep{wang2020minilm}. SciBERT is a BERT-based language model pretrained on a large corpus of scientific publications, generating contextualized embeddings that capture scientific terminology and writing conventions more effectively than general-purpose language models \citep{beltagy2019scibert}. Finally, SPECTER2 is a document-level transformer model specifically designed for scientific publications. Through citation-aware pretraining, it learns representations that incorporate both textual semantics and relationships within the scientific citation graph, producing 768-dimensional embeddings optimized for scientific document understanding \citep{singh2023specter2}. For all embedding-based methods, publication titles and abstracts were concatenated into a single document prior to encoding. Together, these representations provide increasingly rich lexical and semantic descriptions of scientific publications, allowing us to evaluate whether contextual and citation-aware embeddings improve the identification of patent-paper-paired publications relative to traditional lexical approaches.

\textbf{Classifier and hyperparameters.}
Holding the representation fixed on the chosen scientific embedding (SPECTER2), we compared five classifiers spanning probabilistic, linear, and tree-based ensemble families (Table~\ref{table:tab2}, bottom). Na\"ive Bayes is a probabilistic classifier that applies Bayes' theorem under the assumption that features are conditionally independent given the class; despite this simplification it is fast to train and often effective on high-dimensional sparse features \citep{mitchell1997machine}. Logistic regression is a linear model that estimates the probability of the patent-paper-paired class by passing a weighted sum of the input features through the logistic function, providing an interpretable baseline for both sparse and dense inputs \citep{cox1958regression}. Random Forest is an ensemble of decision trees, each grown on a bootstrap sample with a random subset of features considered at each split, whose averaged predictions reduce variance and capture nonlinear interactions \citep{breiman2001random}. Gradient Boosting builds trees sequentially, fitting each new tree to the residual errors of the current ensemble to model complex nonlinear structure \citep{friedman2001gradient}. XGBoost is a regularized, scalable implementation of gradient-boosted trees with efficient training and explicit regularization to limit overfitting \citep{chen2016xgboost}. 

Hyperparameters were tuned with \texttt{RandomizedSearchCV} under five-fold cross-validation, optimizing ROC-AUC; the XGBoost search space included the number of trees ($\mathrm{n\_estimators} = 150$--$1000$), learning rate ($\mathrm{learning\_rate} = 0.01$--$0.60$), subsampling ratio ($\mathrm{subsample} = 0.30$--$0.90$), maximum tree depth ($\mathrm{max\_depth} = 3$--$9$), column sampling ratio ($\mathrm{colsample\_bytree} = 0.50$--$0.90$), and minimum child weight ($\mathrm{min\_child\_weight} = 1$--$4$). All models used the binary logistic objective. Model performance was assessed using multiple evaluation metrics, including receiver operating characteristic area under the curve (ROC-AUC), precision-recall area under the curve (PR-AUC), and confusion matrices. The classifier comparison is shown in Figure~\ref{fig:fig2}.

\textbf{Index construction.}
We used the predicted probability from the optimized classifier as the TRI score. The score ranges from 0 to 1. A higher value means that a publication is more similar to publications in the patent-paper-paired class. A lower value means that the publication is more similar to the matched comparison class. We define the predicted probability of the patent-paper-paired class as the Translation Readiness Index.

Formally, for a publication (i), the Translation Readiness Index is defined as:

\begin{equation}
TRI_i = P(y_i = 1 \mid \mathbf{x}_i),
\end{equation}

where $\mathbf{x}_i$ is the 768-dimensional SPECTER2 embedding for publication $i$, $y_i = 1$ denotes membership in the patent-paper-paired class, and $P(y_i = 1 \mid \mathbf{x}_i)$ is the classifier-estimated probability. TRI can be interpreted as a text-based estimate of similarity to patent-paper-paired science.

TRI measures similarity to patent-paper-paired publications. It does not measure licensing revenue, startup formation, patent value, clinical adoption, or realized commercial success. It also does not imply that a publication will later be patented or cited by a patent. The score is a text-based estimate of proximity to science represented in patent documents.

\section{Results}

\subsection{Linguistic Patterns in Patent-Paper-Paired and Matched Comparison Publications}
The linguistic analysis shows differences between patent-paper-paired publications and matched comparison publications. Table~\ref{table:tab3} reports the main term groups.

A distinctive feature of patent-paper-paired publications is the use of invention-oriented language. A key finding of this research is that there is evidence for the existence of a “language of invention” and that it can be used as a latent feature to distinguish translation-ready research. This is visible in the
over-representation of terms such as \emph{`novel'}, \emph{`propose'}, \emph{`developed'}, \emph{`prototype'}, \emph{`technique'}, and \emph{`design'}, which point to creation, implementation, and technical development. These expressions are consistent with a focus on methods, systems, and applications.

Patent-paper-paired publications also use more language associated with engineered systems, technical artifacts, and implementation. Terms related to hardware and electronics, including \emph{`device'}, \emph{`hardware'}, \emph{`chip'}, \emph{`sensor'}, \emph{`antenna'}, and \emph{`optical'}, appear among the distinguishing features. Signal-processing terms such as \emph{`coding'}, \emph{`compression'}, \emph{`frequency'}, and \emph{`resolution'} are also more common in the patent-paper-paired group.

Matched comparison publications show a different pattern and use more observational and interpretive language. Terms such as \emph{`investigated'}, \emph{`revealed'}, \emph{`obtained'}, \emph{`showed'}, and \emph{`suggest'} are associated with reporting, explanation, and interpretation. Terms associated with observational studies such as those in population health including  \emph{`population'}, \emph{`mortality'}, \emph{`risk'}, and \emph{`health'}, are also more common in the matched comparison group. And some basic scientific terms such as \emph{`carbon'}, \emph{`oxygen'}, \emph{`pathway'}, and \emph{`microbiota'} are also more common in the matched comparison group. These results indicate that despite selecting articles from the same journals, fields and years, the two classes clearly differ linguistically and semantically. The distinction is consistent with the use of patent-paper pairs as a signal of translational orientation.

\begin{table*}[ht]
\centering
\caption{\textbf{Key linguistic themes associated with patent-paper-paired and matched comparison publications.} Thematic categories, descriptions, and representative terms distinguishing patent-like from non-patent-like papers. Patent-paper-paired publications use more language associated with applied technologies, engineered systems, and invention-oriented framing. Matched comparison publications use more language associated with observational research, fundamental science, natural phenomena, and interpretive academic framing. Representative terms are those with the largest frequency and TF-IDF differences between the two classes. They are used for interpretation only and are not inputs to the TRI model. }
  \label{table:tab3}
  \vspace{0.3cm}
  \small
  \makebox[\textwidth][c]{%
  \hspace*{-0.3cm}
  \begin{tabular}{cccc}
    \toprule
    &\textbf{Themes}&\textbf{Description}&\textbf{Representative Terms}\\
    \midrule
    \multirow{4}{*}{\rotatebox[origin=c]{90}{\parbox[c]{4cm}{\centering \textbf{Patent-Like Papers}}}} & \multicolumn{1}{c}{\makecell{Invention \\ metalanguage}}  & \multicolumn{1}{c}{\parbox{6cm}{Action-oriented language around creating, doing and implementing.}} & \multicolumn{1}{c}{\parbox{6cm}{novel, propose, developed, prototype, technique, approach, enables, allows, design}}  \\
    \cmidrule(lr){2-4} 
    & \multicolumn{1}{c}{\makecell{Signal Processing \\\& Systems}}  & \multicolumn{1}{c}{\parbox{6cm}{Engineered signals and system operations.}} & \multicolumn{1}{c}{\parbox{6cm}{signal, coding, decoding, compression, noise, frequency, resolution, pulse}} \\
    \cmidrule(lr){2-4} 
    & \multicolumn{1}{c}{\makecell{Applied Biotechnology \\(Therapeutics)}}  & \multicolumn{1}{c}{\parbox{6cm}{Engineered biological applications.}} & \multicolumn{1}{c}{\parbox{6cm}{antibody, vaccine, peptide, antigen, engineered}} \\
    \cmidrule(lr){2-4} 
    & \multicolumn{1}{c}{\makecell{Hard Technology \\\& Electronics}}  & \multicolumn{1}{c}{\parbox{6cm}{Physical devices and engineered components.}} & \multicolumn{1}{c}{\parbox{6cm}{device, hardware, chip, memory, architecture, antenna, radar, voltage, camera, sensor, imaging, optical}} \\

    \midrule
    \multirow{4}{*}{\rotatebox[origin=c]{90}{\parbox[c]{4cm}{\centering \textbf{Non-Patent-Like Papers}}}} & \multicolumn{1}{c}{\makecell{Academic \\ metalanguage}}  & \multicolumn{1}{c}{\parbox{6cm}{Observation-focused language focused on analysis and description.}} & \multicolumn{1}{c}{\parbox{6cm}{investigated, obtained, revealed, showed, suggest, considered}} \\
    \cmidrule(lr){2-4} 
    & \multicolumn{1}{c}{\makecell{Fundamental Chemistry \\\& Biology}}  & \multicolumn{1}{c}{\parbox{6cm}{Basic elements and biological systems.}} & \multicolumn{1}{c}{\parbox{6cm}{carbon, hydrogen, oxygen, metal, microbiota, macrophages, pathway}} \\
    \cmidrule(lr){2-4} 
    & \multicolumn{1}{c}{\makecell{Natural Processes \& \\Abstract Phenomena}}  & \multicolumn{1}{c}{\parbox{6cm}{Natural laws and theoretical processes.}} & \multicolumn{1}{c}{\parbox{6cm}{reaction, formation, diffusion, interaction, transition, evolution, dynamics}} \\
    \cmidrule(lr){2-4} 
    & \multicolumn{1}{c}{\makecell{Clinical \&\\ Observational Health}}  & \multicolumn{1}{c}{\parbox{6cm}{Population-level and observational studies.}} & \multicolumn{1}{c}{\parbox{6cm}{study, patients, population, mortality, risk, disease, health, COVID-19}} \\
  \bottomrule
  \end{tabular}}
\end{table*}

\clearpage

\subsection{Model Framework for TRI} \label{baseline_models}

\subsubsection{Classification Performance}
\begin{figure}[ht]
  \centering
  \hspace*{-1.5cm}
  \includegraphics[width=1.2\textwidth]{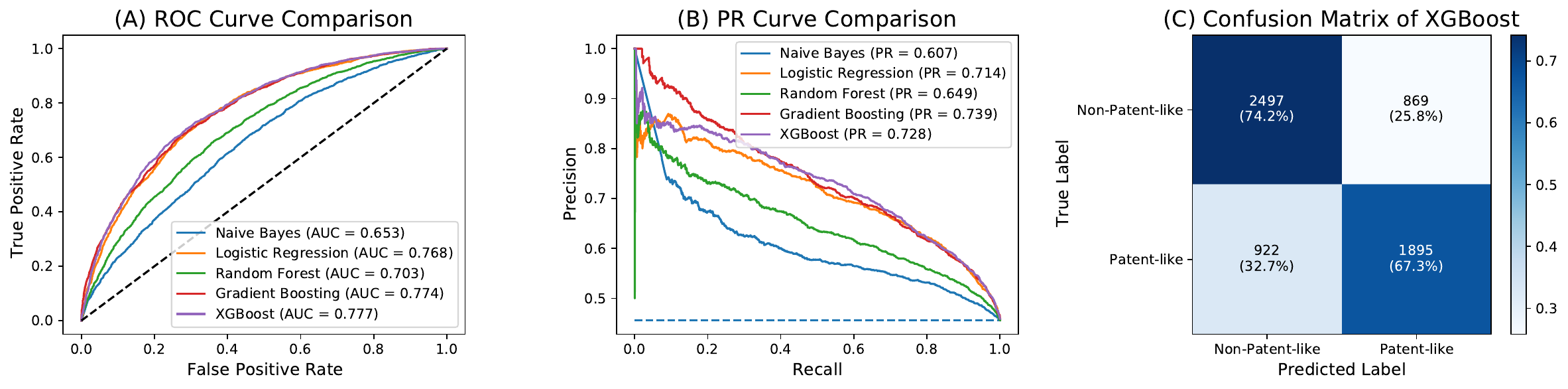}
  \vspace{0.1pt}
  \caption{\textbf{Classification performance for identifying patent-paper-paired publications.} (A) ROC curve comparison and (B) Precision–Recall curve comparison for five machine learning classifiers. XGBoost achieved the best overall ROC performance (AUC = 0.777), while Gradient Boosting achieved the highest precision–recall performance (PR-AUC = 0.739). (C) Confusion matrix of the selected XGBoost classifier, indicating correct classification rates of 74.2\% for non-patent-like papers and 67.3\% for patent-like papers. The results show that semantic embeddings can distinguish patent-paper-paired publications from matched comparison publications with moderate accuracy.}
  \label{fig:fig2}
\end{figure}

Figure~\ref{fig:fig2} compares five classifiers for distinguishing patent-paper-paired publications from matched comparison publications.

With tuning, XGBoost achieved the highest ROC-AUC among the evaluated classifiers (0.777), marginally ahead of Gradient Boosting (0.774), which achieved the highest precision–recall area under the curve (PR-AUC = 0.739). Logistic Regression, Random Forest, and Naïve Bayes performed substantially less well. Without tuning, the same SPECTER2 + XGBoost combination reaches 0.774 (Figure 3); we report this more conservative figure as the headline performance of TRI.

Although the difference between XGBoost and Gradient Boosting is small, XGBoost was selected because it had the highest ROC-AUC. The results should not be read as evidence that XGBoost is uniquely suited to this task. We use XGBoost probabilities to construct TRI because the model provides a continuous score for each publication.

The confusion matrix reports class-specific performance. The model correctly identified 74.2\% of matched comparison publications and 67.3\% of patent-paper-paired publications. These results indicate that title and abstract text and related semantic embeddings contain information associated with the patent-paper-paired class.

The objective of TRI is not only binary classification. Rather, the classifier provides a continuous probability distribution that reflects varying degrees of translational orientation. TRI should be interpreted as a continuous similarity score rather than a categorical label.

\subsubsection{Model Combination Comparison}
\begin{figure}[ht]
  \centering
  \includegraphics[width=0.6\textwidth]{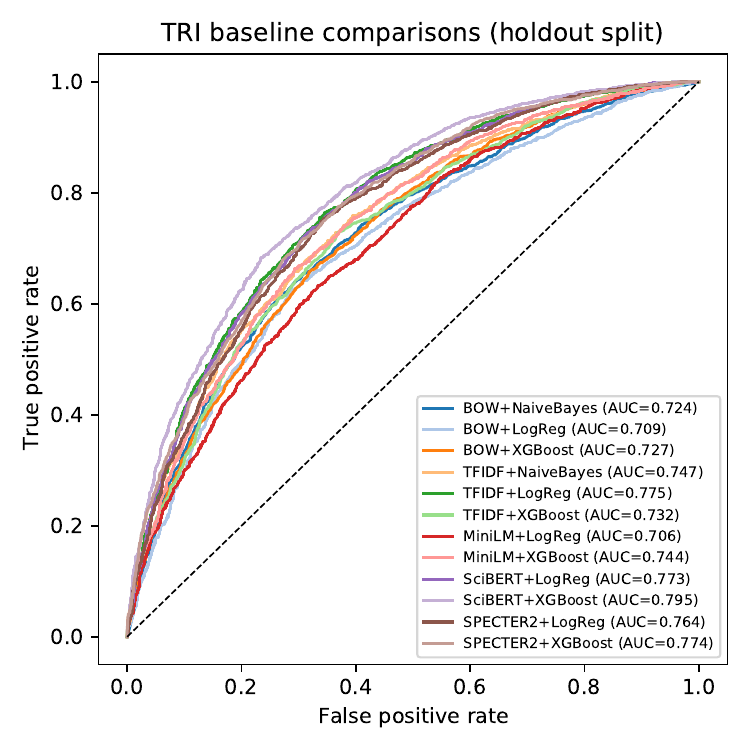}
  \vspace{0.1pt}
  \caption{\textbf{Model combination comparison for TRI.} Held-out ROC curves for five text representations (bag-of-words, TF-IDF, MiniLM sentence embeddings, SciBERT, and SPECTER2) crossed with Na\"ive Bayes, logistic regression, and XGBoost classifier, for the twelve combinations shown. All combinations use default classifier hyperparameters for comparability; Figure 2 reports tuned performance. SciBERT $+$ XGBoost achieved the highest ROC-AUC (0.795); TF-IDF $+$ logistic regression (0.775) and SPECTER2 $+$ XGBoost (0.774) were close behind. The clustering of the strongest scientific-text models indicates a robust underlying signal rather than an artifact of any single model.}
  \label{fig:fig3}
\end{figure}

We use the ROC-AUC curve to compare different model combinations, summarized in Figure~\ref{fig:fig3}. To keep the twelve combinations comparable, the benchmark in Figure 3 uses default hyperparameters for every classifier; the tuned results reported in Section 4.2.1 and Figure 2 are therefore marginally higher for the same representation. The comprehensive evaluation identified the combination of SPECTER2 and XGBoost as the most appropriate framework for constructing the Translation Readiness Index (TRI). The strongest single combination came from a scientific-domain embedding model (SciBERT + XGBoost, ROC-AUC = 0.795), and both scientific embedding models outperformed the general-purpose MiniLM encoder. While lexical approaches such as bag-of-words and TF-IDF capture differences in term occurrence and document-specific vocabulary, transformer-based representations additionally encode contextual and semantic relationships between scientific concepts. The clustering of the leading combinations within a narrow band indicates that the underlying signal is recoverable by several distinct representations rather than being an artifact of any one.

Although the combination of SciBERT and XGBoost achieved the highest ROC-AUC in the benchmark comparison, the improvement over SPECTER2-XGBoost was modest. Rather than selecting the highest-scoring model solely on predictive performance, the choice of representation was guided by the intended purpose of TRI as a reproducible, scalable measurement framework. SPECTER2 is specifically designed to generate document-level embeddings for scientific publications using citation-aware pretraining, allowing it to capture both the semantic content of a publication and its broader position within the scientific knowledge network. Because it produces a single representation from the title and abstract of any publication, it can be applied consistently across millions of papers without requiring citation features at inference time. Furthermore, SPECTER2 has become a widely adopted representation in science-of-science and bibliometric research, making TRI more readily comparable with existing studies and facilitating future replication and extension.

The classifier comparison similarly supported the use of XGBoost. Among the evaluated classification algorithms, XGBoost achieved the highest ROC-AUC while consistently demonstrating strong discrimination between patent-paper-paired and matched comparison publications. Its gradient-boosting framework effectively captures nonlinear relationships and complex interactions within high-dimensional embedding spaces, making it well suited to document-level representations such as SPECTER2. In addition, the regularization mechanisms incorporated into XGBoost reduce the risk of overfitting while maintaining computational efficiency, making it an appropriate choice for large-scale inference.

Accordingly, TRI is defined as the predicted probability generated by an XGBoost classifier operating on SPECTER2 document embeddings. Importantly, the relatively small differences observed among the highest-performing scientific embedding models indicate that the detected signal is not dependent on a particular embedding architecture. Instead, the results suggest that translational orientation is a robust semantic characteristic encoded within the language of scientific publications, and that modern scientific document embeddings are consistently able to recover this signal. This finding increases confidence that TRI measures an underlying property of publication text rather than an artifact of a specific representation or classifier choice.


\subsection{Validation of TRI: UWA Publications}
We applied TRI to 25,921 publications authored by researchers affiliated with the University of Western Australia between 2019 and 2026. This corpus was not used for model training. The validation analyses test whether higher TRI scores are associated with external indicators related to research translation. These analyses do not show that TRI measures commercialization. They test whether TRI is directionally related to translation-adjacent measures.

\subsubsection{Commercial-Potential Score Comparison}
The first validation examined the relationship between TRI and publication-level commercial potential scores obtained from the ComPot dataset. Commercial-potential scores provide an external comparison measure related to possible commercial or technological value. Specifically, they are estimated using large language models and deep neural networks that analyze the publication's abstract to predict the probability that it will subsequently be cited in a renewed patent, thereby providing an ex-ante estimate of its commercial potential, independent of realized commercialization outcomes. To compare the two measures, publications were grouped by increasing TRI thresholds, and the average commercial-potential score was calculated at each threshold.

As shown in Figure~\ref{fig:fig4} (A), average commercial-potential scores increased at higher TRI thresholds. Publications with relatively low TRI values exhibited average commercial potential scores below 0.50, whereas publications with TRI values approaching 1.0 achieved average scores exceeding 0.70. The increase was larger among the highest-TRI publications. The upward trend suggests that publications with higher TRI scores also tend to receive higher commercial-potential scores in this validation set. This supports the interpretation of TRI as a translation-oriented text measure, although the relationship should be interpreted with caution as both measures use related textual and patent-based signals albeit with different approaches.

\subsubsection{Pasteur’s Quadrant Researcher Comparison}
The second comparison used the Pasteur’s Quadrant Research Score dataset (\cite{pqrs2024}), which identifies researchers who are “inventor-scholars” or people who publish papers and patents and thus are associated with use-inspired research. Pasteur’s Quadrant is commonly used to describe research that combines basic research questions with practical use.

To compare TRI with this measure, we counted PQRS researchers associated with publications above increasing TRI thresholds. Two complementary measures were evaluated: (1) whether the first author appeared in the PQRS database, and (2) whether any coauthor appeared in the PQRS database.

Figure~\ref{fig:fig4} (B) shows that publications with higher TRI values more often include researchers identified in the PQRS data. While the number of first authors appearing in the PQRS dataset remained relatively modest across thresholds, a substantially stronger pattern emerged when considering all coauthors. The number of PQRS-affiliated coauthors increased among publications with TRI values above 0.90. This result shows that high-TRI publications more often include researchers classified in the PQRS data. The result is consistent with the intended interpretation of TRI.

\subsubsection{Comparison of Top-, Middle-, Bottom-, and Random-Ranked 100 UWA Publications by TRI}

To assess whether higher TRI scores are associated with established indicators of translational activity, we compared four groups of 100 UWA publications. The Top 100 and Bottom 100 comprised the publications with the highest and lowest TRI scores, respectively. The Middle 100 were randomly sampled from publications with TRI scores between 0.47 and 0.53, which is close to the TRI threshold. The Random 100 were sampled from the remaining publications after excluding the Top, Middle, and Bottom groups.

We examined three indicators: PQRS researcher involvement, UWA-related patent-author involvement, and company coauthorship. For the patent indicator, we restricted Google Patents to 609 patents related to UWA and matched their associated authors to publication authors. As shown in Figure~\ref{fig:fig4}(C), the Top 100 had the highest prevalence of all three indicators. PQRS involvement was observed in 55\% of the Top 100, compared with 44\%, 17\%, and 34\% in the Middle, Bottom, and Random groups, respectively. Patent-author involvement showed an even stronger gradient, declining from 37\% in the Top 100 to 18\% in the Middle 100, 1\% in the Bottom 100, and 9\% in the Random 100. Company coauthorship was less frequent overall, occurring in 9\% of the Top 100, compared with 5\%, 0\%, and 5\% in the other groups.

Relative to the Random 100, Top-100 publications had significantly higher odds of PQRS involvement ($OR = 2.37$, $95$\% $CI: 1.34-4.20$; Fisher's exact ($p=0.004$)) and UWA-related patent-author involvement (OR = $5.94$, $95$\% $CI: 2.68-13.17$; ($p<0.001$)). The odds of company coauthorship were also higher ($OR = 1.88$), although this difference was not statistically significant ($p=0.407$). The marked decline in patent- and PQRS-related involvement from the Top to the Middle and Bottom groups further indicates a systematic gradient across TRI levels. Overall, these results provide external validation that higher TRI scores are associated with research embedded in stronger translational and commercial contexts.

\subsubsection{Institutional Coauthorship Comparison}
Finally, we examined organizations coauthoring high-TRI UWA publications. For each external organization coauthoring high-TRI UWA publications, TRI scores were aggregated at the institutional level.

\begin{figure}[ht]
  \centering
  \hspace*{-1.5cm}
  \includegraphics[width=1.2\textwidth]{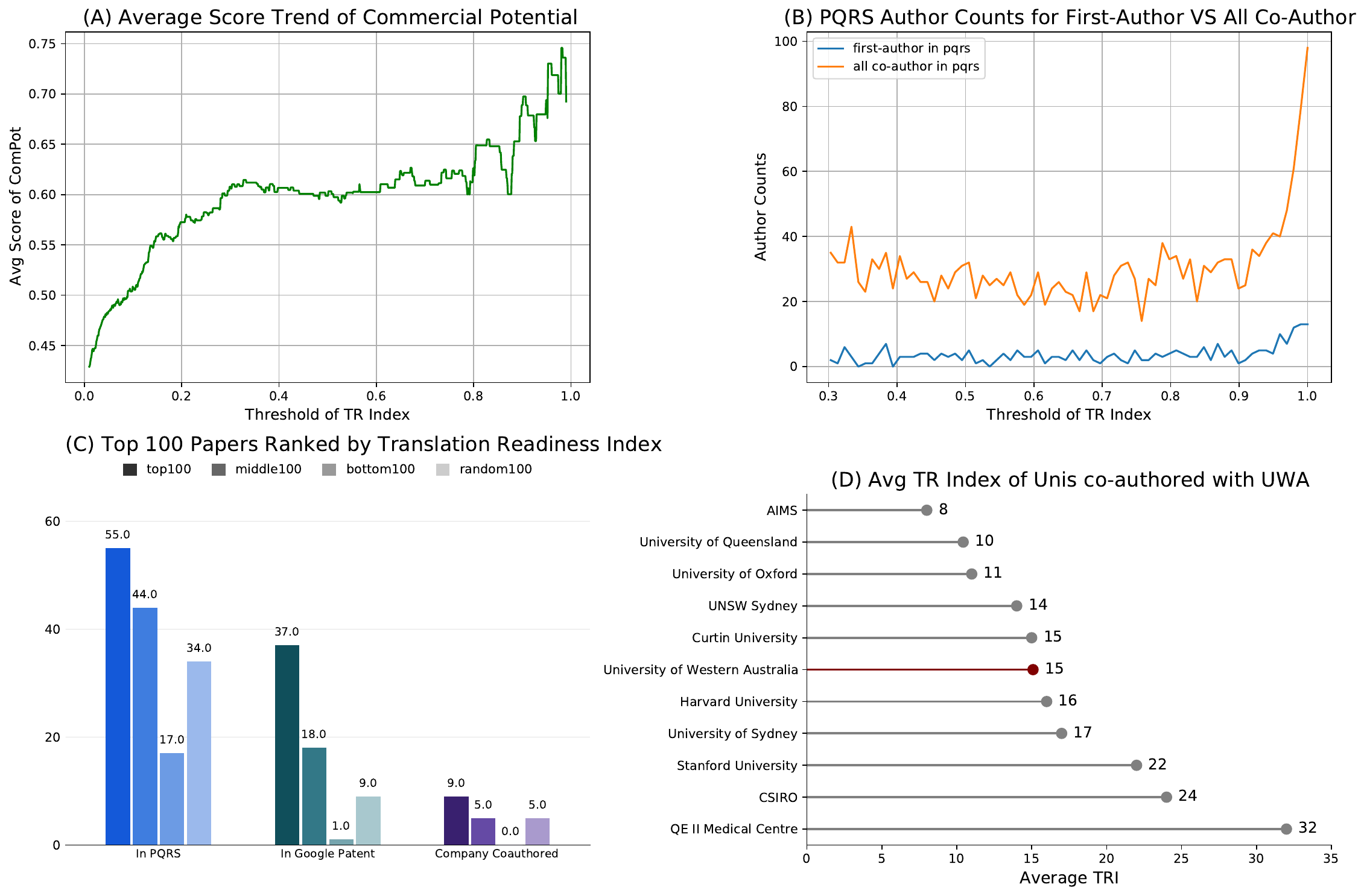}
  \vspace{0.1pt}
  \caption{\textbf{External validation of the Translation Readiness (TR) Index.} (A) Average commercial potential scores across increasing TR Index thresholds. (B) Counts of Pasteur Quadrant-affiliated authors among papers exceeding different TR Index thresholds. (C) Comparison among Top-, Middle-, Bottom-, Random- 100 UWA publications by TR Index. Using randomly selected publications as a benchmark, Top-100 papers had 2.4x higher odds of PQRS involvement, 5.9x higher odds of UWA patent-author involvement, and 1.9x higher odds of company coauthorship. (D) Average TR Index of institutions coauthoring with UWA, reported on a 0–100 scale. The results provide external checks on whether TRI is associated with indicators related to research translation. These associations support the interpretation of TRI as a measure of translational orientation.}
  \label{fig:fig4}
\end{figure}

Figure~\ref{fig:fig4} (D) shows the external organizations most frequently appearing in high-TRI UWA coauthored publications. Among high-TRI UWA publications, the most frequent external collaborators were QE II Medical Centre (32 publications), CSIRO (24), Stanford University (22), the University of Sydney (17), Harvard University (16), and the University of Oxford (11). Several Australian research-intensive universities, including UNSW Sydney, Curtin University, and the University of Queensland, also appeared prominently among the highest-ranking collaborators.

These institutions are characterized by strong records in technology transfer, industry collaboration, biomedical innovation, and commercialization activities. Their presence among high-TRI UWA publications is consistent with the interpretation that TRI captures text associated with research translation. This result should be interpreted descriptively, since collaboration patterns also reflect geography, field mix, and institutional networks. The four validation analyses are consistent with the interpretation of TRI as a publication-level measure of translational orientation. They do not establish that TRI measures realized commercialization.

\clearpage

\subsection{Validation of TRI: Institutional Industry Collaboration}
To evaluate TRI at the institutional level, we constructed a dataset of publications from the 2025 Aggregate Ranking of Top Universities (ARTU) Top 100 institutions. Each university was matched to its corresponding OpenAlex institution identifier, with manual verification performed where name disambiguation was required. We then retrieved all peer-reviewed journal articles published in 2025 for which the institution appeared in the authorship lineage. To focus on internationally recognized research while reducing disciplinary differences in citation practices, we restricted the corpus to publications ranked in the top 1\% of their field by normalized citation percentile. Publication metadata, including titles, abstracts, authorship, and institutional affiliations, were collected, and only publications with complete title and abstract information were retained for TRI inference. Institution-level TRI was computed by aggregating publication-level scores. To ensure reliable institutional estimates, we included only universities with at least 30 eligible publications, resulting in a final sample of 92 universities.

\subsubsection{TRI Representative for Institutions}
To compare translational orientation across universities, publication-level TRI scores must be aggregated into a single institutional statistic. Because TRI is a continuous publication-level measure, several summary statistics are plausible, including the arithmetic mean, median, upper quantiles, and the proportion of publications exceeding predefined TRI thresholds. However, these alternatives differ in their statistical reliability and their ability to distinguish institutions. An appropriate institutional representative should exhibit substantial variation between universities while remaining stable with respect to sampling variability within each university.

We therefore evaluated five candidate summary statistics: the arithmetic mean, median, the proportion of publications with TRI greater than 0.5, the proportion with TRI greater than 0.9, and the 90th percentile of the TRI distribution. For each university, uncertainty in each statistic was estimated via bootstrap resampling of publications, thereby decomposing the observed variation into between-university and within-university sampling variation. Reliability was then defined as the proportion of the total variance attributable to genuine differences between universities rather than estimation noise.

To evaluate the reliability of each candidate institutional summary statistic, we decomposed the observed between-university variation into signal and estimation noise. Reliability was defined as

\begin{equation}
R =
\frac{\sigma_{\mathrm{between}}^{2}-\overline{\sigma_{\mathrm{within}}^{2}}}
{\sigma_{\mathrm{between}}^{2}},
\end{equation}

where $\sigma_{\mathrm{between}}^{2}$ is the variance of the estimated statistic across universities and $\overline{\sigma_{\mathrm{within}}^{2}}$ is the mean bootstrap variance of the statistic within universities. This quantity estimates the proportion of the observed variance attributable to genuine differences between universities rather than sampling variability. Higher values therefore indicate that a statistic more reliably distinguishes institutions.

\begin{figure}[ht]
  \centering
  \includegraphics[width=0.8\textwidth]{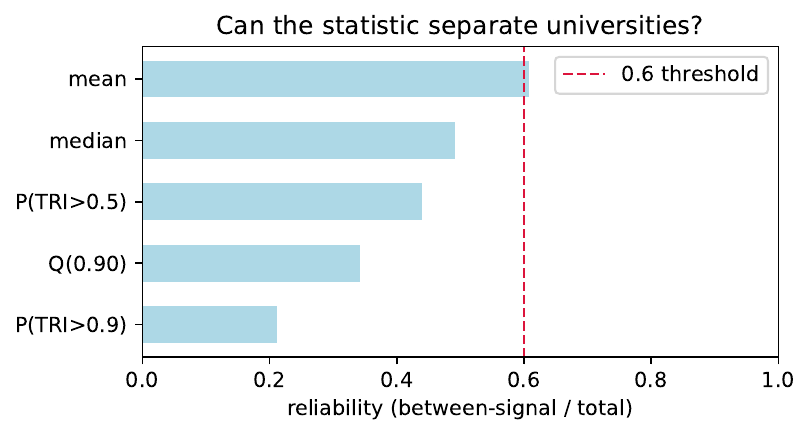}
  \vspace{0.1pt}
  \caption{\textbf{Reliability of candidate institution-level TRI summary statistics.} Reliability is defined as the proportion of total variance explained by true between-university differences after accounting for within-university sampling variability. Among the evaluated statistics, the arithmetic mean achieved the highest reliability, indicating that it best separates universities while remaining robust to sampling variation. The dashed line indicates a reliability threshold of 0.6.}
  \label{fig:fig5}
\end{figure}

Figure~\ref{fig:fig5} summarizes the resulting reliability estimates. The arithmetic mean achieved the highest reliability among all candidate statistics, indicating that it provides the clearest separation between institutions while remaining the most stable under repeated sampling. In contrast, statistics based on the upper tail of the TRI distribution, such as the proportion of publications with TRI greater than 0.9 and the 90th percentile, exhibited substantially lower reliability because they depend on relatively few publications and are therefore more sensitive to sampling variability. Although the median is robust to extreme observations, it also showed lower reliability than the mean, reflecting the concentration of many institutional TRI distributions around similar central values. Threshold-based measures similarly discarded much of the information contained in the continuous TRI scores, reducing their ability to discriminate among universities.

These findings indicate that the arithmetic mean provides the most appropriate institution-level representation of TRI. By incorporating information from the full distribution of the publication, the mean achieves the best balance between statistical stability and discriminatory power. Consequently, all subsequent institution-level analyses use the mean TRI as the institutional summary statistic. This choice also facilitates interpretation, as the mean represents the expected translational orientation of publications produced by a university while making efficient use of the TRI score's continuous probabilistic nature.

\subsubsection{Correlation with Industry Collaboration}
We compared institutional mean TRI scores with industry collaboration indicators for universities in the ARTU Top 100. This comparison tests whether institutions with higher average TRI scores also have higher industry collaboration scores.

For each university, publication-level TRI scores were averaged to produce an institutional TRI score. These values were compared with CWTS Leiden Ranking industry collaboration indicators, which measure the share of publications coauthored with industrial partners.

\begin{figure}[ht]
  \centering
  \includegraphics[width=0.9\textwidth]{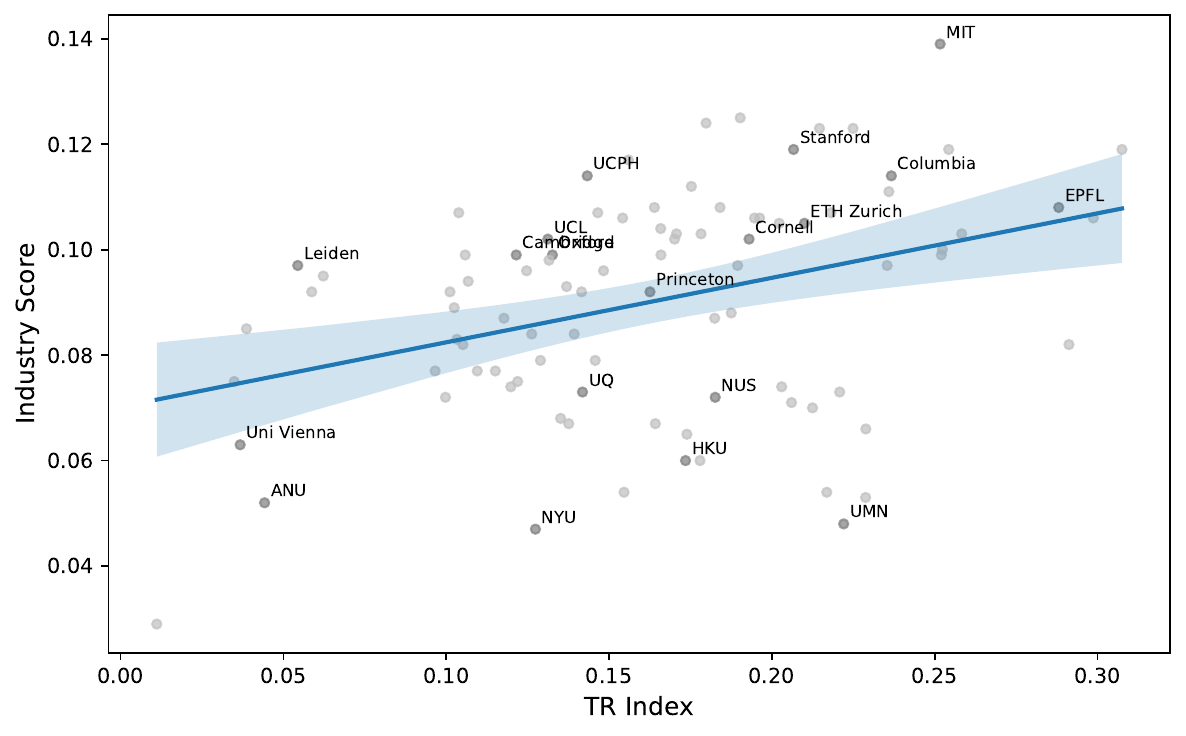}
  \vspace{0.1pt}
  \caption{\textbf{Association between the Translation Readiness (TR) Index and university industry engagement.} Relationship between the average TR Index and industry engagement scores for universities ranked in the ARTU Top 100. The fitted regression line and 95\% confidence interval show a positive but modest association between average TRI and industry collaboration.}
  \label{fig:fig6}
\end{figure}

Figure~\ref{fig:fig6} shows the relationship between average institutional TRI scores and industry collaboration scores. The two measures are positively associated, although institutions differ in mission, field mix, geography, and national context. Universities with higher average TRI scores generally exhibit stronger industry collaboration performance, whereas institutions with lower TRI values tend to demonstrate weaker engagement with industrial partners. MIT, Stanford University, ETH Zurich, and EPFL have relatively high values on both measures. Institutions with lower TRI values tend to have lower industry collaboration scores.

We estimated a linear regression with industry collaboration score as the dependent variable and average TRI as the independent variable. The association was positive and statistically significant: $r = 0.364, \beta = 0.122, p < 0.001, R^2 = 0.133$. The effect is modest. Average TRI explains about 13.3\% of the variation in the industry collaboration measure. This is consistent with TRI capturing one component of institutional translational orientation, while most variation is likely explained by field mix, institutional strategy, national context, and local industry structure. Also, we can see too that by looking at peers one could establish a benchmark range of expected TRI for a university with a given level of industry engagement.  Those with above average TRI relative to their peers could perhaps have latent potential for greater involvement.

TRI is derived from publication title and abstract text and does not include patents, industrial partnerships, licensing activity, institutional rankings, or commercialization outcomes as model inputs. The observed relationship supports the interpretation that publication text is associated with some external indicators of research translation. The institutional results provide limited but useful external validation. Universities with higher average TRI scores tend to have higher industry collaboration scores, but the relationship is weak. TRI should not be treated as a substitute for direct measures of university-industry collaboration.

\clearpage

\section{Discussion}
This paper tests whether publication text alone can be used to distinguish research appearing in patent-paper pairs from matched comparison publications using large-scale semantic embeddings. Many indicators used to study research translation, including patents, licensing revenue, startup formation, and industry collaboration, are observed after publication and are affected by institutional context. Universities, funders, investors, and technology-transfer offices therefore often need earlier indicators of translational orientation. This study introduces TRI, a text-based score estimating similarity to publications in high-confidence patent-paper pairs.

Using more than 20,000 publications, we show that patent-paper-paired publications differ from matched comparison publications and that difference relates in part to a “language of invention” revealed in the title and abstract text. Our selected model combination reached a ROC-AUC of 0.774 on the held-out test set. The results show measurable language differences between patent-paper-paired publications and matched comparison publications. Patent-paper-paired publications more often use invention-oriented framing and terms linked to devices, engineered systems, signal processing, and applied biotechnology. Matched comparison publications more often use clinical, observational, and explanatory language. These findings show that language differs between publications represented in patent-paper pairs and matched comparison publications.

TRI was compared with several external indicators. In the UWA validation set, higher TRI scores were associated with commercial-potential scores, patent-author involvement, PQRS coauthorship, and industry coauthorship. At the institutional level, average TRI had a positive but modest association with industry collaboration. TRI is derived from title and abstract text and does not use patents, licensing activity, institutional rankings, or commercialization outcomes as model inputs. These results support the use of TRI as a text-based measure of translational orientation. They do not show that TRI measures commercial success.

The data science contribution is the measurement pipeline. Citation-based indicators measure later attention to a publication. TRI uses title and abstract text to estimate similarity to patent-paper-paired science. The study links publication metadata, patent-paper-pair labels, scientific document embeddings, supervised classification, and external validation. The pipeline can be applied to other definitions of research translation if suitable labels are available.

The study has several limitations. First, the positive class is based on patent-paper pairs. This captures one route by which science enters the patent system, but it does not capture all forms of social, clinical, environmental, policy, or open-source impact. Important forms of impact, including policy influence, clinical implementation, environmental outcomes, and open-source innovation, are not directly represented in the current model. Second, the comparison class consists of publications not observed in the patent-paper-pair corpus. Some of these publications may still have translational value. Third, TRI may reflect field differences even after matching by year, venue, and subfield. Fourth, the model uses only title and abstract text. Full text, funding information, author history, institutional context, and citation networks may change the results. TRI should therefore be used as a screening measure, not as a decision rule.

Future work could test other impact labels, add full-text data, and compare TRI with field-specific indicators. The measure could be adapted to other forms of research use if suitable labels are available. Other inputs, including citation networks, funding information, institutional characteristics, and full text, could change model performance and interpretation. Future studies could also compare TRI across careers, research groups, institutions, and technology areas.

TRI should not be used on its own for funding, hiring, promotion, or technology-transfer decisions. The score may penalize fields where translation occurs through clinical guidelines, public policy, software, standards, community practice, or environmental management rather than patents. It may also reflect language style, discipline, institution, and geography. TRI is intended to support discovery and prioritization, not to replace expert review.

This study introduces TRI as a publication-level measure of similarity to patent-paper-paired science. The method combines patent-paper-pair labels, semantic paper-level embeddings, and supervised classification. The results show that title and abstract text and related semantic embeddings contain information associated with patent-paper-paired science. TRI may help users screen publications whose language resembles science represented in the patent system. It should be used with expert review and field-specific context.

\subsection*{Data Availability}
Publication metadata used in this study were obtained from OpenAlex. Patent-paper-pair data were obtained from the Reliance on Science patent-paper-pairs dataset. Pasteur's Quadrant classifications were obtained from the Reliance on Science Pasteur's Quadrant Researchers dataset. Patent-author information was derived from Google Patents. Institutional indicators were derived from ARTU and the CWTS Leiden Ranking. OpenAlex metadata can be reconstructed from publication identifiers and query logic. RoS, PQRS, ComPot, ARTU, CWTS Leiden Ranking, and Google Patents data are subject to the terms of their respective providers. Where redistribution is not permitted, source identifiers and processing steps are provided where allowed.

\subsection*{Code Availability}
All analyses were conducted using Python. Major software libraries included SPECTER2, Scikit-learn, XGBoost, Pandas, NumPy, and Matplotlib. Analysis code, model-training scripts, matching code, and figure-generation scripts will be made available in a public repository.

\subsection*{Funding}
This research received no specific grant from any funding agency in the public, commercial, or not-for-profit sectors.

\subsection*{Disclosure Statement}
The authors have no conflicts of interest to declare.

\subsection*{Acknowledgments}
The authors acknowledge OpenAlex, Google Patents, ARTU, CWTS Leiden Ranking, and Reliance on Science as data sources used in this study.
 
\subsection*{Contributions}
P.X.M. conceived the study. X.G. and R.A. collected and processed the data, implemented the machine learning framework, and conducted the analyses. All authors contributed to the interpretation of results and manuscript preparation.

\clearpage




\printbibliography

\end{document}